\documentclass[12pt]{article}
\usepackage{amsmath}
\usepackage{graphicx,psfrag,epsf}
\usepackage{enumerate}
\usepackage{natbib}
\usepackage{amsfonts,amssymb,amsmath,amsthm}
\usepackage{bbm}
\usepackage{siunitx}

\newcommand{\blind}{0}

\newcommand{\Rset}{\mathbb{R}}								
\newcommand{\Xset}{\mathbb{X}}	
\newcommand{\Nset}{\mathbb{N}}								

\newcommand \trans {^\top}
\newcommand{\esp}{\mathbb{E}}

\begin{document}

\def\spacingset#1{\renewcommand{\baselinestretch}%
{#1}\small\normalsize} \spacingset{1}


\if0\blind
{
  \title{\bf Modeling non-stationary extreme dependence with stationary max-stable processes and multidimensional scaling}
  \author{Cl\'ement Chevalier\thanks{
  The authors would like to warmly thank  
Dr. Sophie Fukutome from Meteo-Swiss for useful scientific discussions 
and for kindly providing the precipitation dataset. 
The authors also thank Prof. Sebastian Engelke  
for the productive discussions and the suggestion to investigate Brown-Resnick models.
}
  \hspace{.2cm}\\
    Institute of Statistics, University of Neuch\^atel\\
    and \\
    Olivia Martius \\
    Institute of Geography, University of Bern\\
    and \\
    David Ginsbourger \\
    Idiap Research Institute, Martigny,  \\
IMSV, 
University of Bern
    }
  \maketitle
} \fi

\if1\blind
{
  \bigskip
  \bigskip
  \bigskip
  \begin{center}
    {\LARGE\bf Title}
\end{center}
  \medskip
} \fi

\bigskip
\begin{abstract}
Modeling the joint distribution of extreme weather events in multiple locations is a challenging task with important applications. In this study, we use max-stable models to study extreme daily precipitation events in Switzerland. 
The non-stationarity of the spatial process at hand involves important challenges, which are often dealt with by using a stationary model in a so-called climate space, with well-chosen covariates.
Here, we instead chose to warp the weather stations under study in a latent space of higher dimension using multidimensional scaling (MDS). The advantage of this approach is its improved flexibility to reproduce highly non-stationary phenomena, while keeping a tractable stationary spatial model in the latent space. 
Two model fitting approaches, which both use MDS, are presented and compared to a classical approach that relies on composite likelihood maximization in a climate space. 
Results suggest that the proposed methods better reproduce the observed extremal coefficients and their complex spatial dependence.
\end{abstract}

\noindent%
{\it Keywords:}  Extremal coefficient, Extreme value theory, Spatial extremes

\spacingset{1.45}
\section{Introduction}
\label{sec:intro}

Understanding the joint distribution of extreme rainfall at different locations is a challenging problem with important stakes. Extreme rainfall can result in damage due to  extensive overland flow and cause natural disasters such as mudslides \citep{RN957} or floods \citep{RN1007}. In the Alpine area, extreme precipitation events are known to be more intense than in other parts of Europe \citep{Frei:1998}. The methods developed in this work are applied to data from Switzerland, where  extreme rainfall and associated flooding can have substantial socio-economic consequences \citep{Hilker:2009}. A significant fraction of extreme daily  rainfall events in Switzerland is associated with synoptic-scale weather systems \citep{Martius:2006,Pfahl:2014,Giannakaki:2015}. These rainfall events typically affect wider areas and are spatially dependent. 

\smallskip

Spatial models for extremes have received particular attention in the last few years, through the use of max-stable process modeling  \citep{de1984spectral,smith1990,schlather2002models,kabluchko2009stationary}.  Although  statistical inference is known to be challenging  \citep{bacro2014:maxstableestimation}, max-stable processes have been used for spatial modeling of extreme temperatures \citep{davison2012geostatistics}, winds \citep{engelke2015JRSSB}, precipitation  \citep{smithstephenson2009,padoanribatetsisson2010,huserdavison2014spacetime,shang2015}  and snow depths \citep{blanchet2011spatial,gaume2013snowfalls}. The theory of max-stable processes generalizes \citep{ribatet2013} the now well-established univariate extreme value theory \citep{coles2001introduction}. Although other approaches exist for constructing spatial models for extremes (see, \citet{davison2012review} for a review), this work focuses on the joint distributions of extremes through max-stable processes.

\smallskip

Max-stable models are often calibrated in a so-called climate space (see, \citet{blanchet2011spatial} and references therein), which is a parametric transformation of the space of spatial coordinates with additional well-chosen covariates.  This approach is used in, e.g., \cite{cooley2007bayesian,blanchet2011spatial} and has been shown to improve the fits compared to models that are merely calibrated in the 2 or 3-dimensional space of spatial coordinates. The use of additional covariates can be seen as a solution to deal with \textit{non-stationarity}. Indeed, the dependence between extremes often cannot be seen as a simple function of the Euclidean distance between the stations in the (longitude, latitude) space or the  (longitude, latitude, elevation) space, even if some anisotropy is added. \citet{Frei:1998} and \citet{Fukutome:2014} have shown that the climatological spatial distribution of both mean and extreme precipitation in Switzerland is not a simple function of height and location because of slope and shielding effects. The orientation of the topography relative to the incoming moist air flow plays a prominent role in the distribution of spatial precipitation. An example event is presented in \citet{Roessler:2014}, where the amount of  precipitation falling on the north- and south-facing slope of a valley differed by a factor of three due to local circulation effects. A natural solution for overcoming non-stationarity problems is to tune distances between stations through additional covariates and parametric space transformation. The limitation here is that there is no guarantee to produce a space in which the observed dependencies are stationary. 

\smallskip

In this work, we also choose to deal with non-stationarity by using a stationary model in a different space. Instead of a climate space, we rely on a higher dimensional latent space in which the different stations under study are warped. We propose two different model fitting approaches, both of which use multidimensional scaling (MDS) to place the stations in the latent space. When the latent space dimension is not too large, the mapping used to warp the stations in the latent space turns out to be smooth enough to allow interpolation. Any location in Switzerland or any station that was not used to fit the model can thus be placed in the latent space. Ultimately, non-stationarity in the (longitude, latitude, elevation) space is  efficiently reproduced using a stationary model in the latent space. In comparison, simple models based on climate spaces fail to reproduce the observed non-stationary dependencies. Of the possible choices for max-stable processes, we chose to focus on  Brown-Resnick models \citep{kabluchko2009stationary,davison2012review}, even though other models have been considered.

\smallskip

Max-stable models and MDS are well-established concepts. The novelty of this work lies in the new max-stable model fitting approach, which relies on MDS. In \textit{non-extreme} spatial statistics, MDS has notably been used by \citet{sampsonguttorp1992JASA} to build a spatio-temporal kriging model. For the spatial part of their model, the dimension of the latent space built with MDS is  set to $2$, allowing the creation of a map that warps any location of the input space to the latent space. More recently, \citet{bornn2012} suggested dealing with non-stationarity by using a stationary model in an expanded space, which they defined as the product of the initial input space and a 1- or 2-dimensional space built with MDS. In the framework of max-stable modeling, MDS appears to be a logical solution for fitting stationary Brown-Resnick models since the pairwise extremal coefficients  between stations (see, Section~\ref{sec:Extremalcoefficients} for a definition) are directly linked to their interdistance. One may thus ``play'' with this  distance and place the stations in the latent space such that the modeled extremal coefficients match the extremal coefficients estimated from the data.  This idea motivated our first model-fitting method. The second model-fitting approach relies on both MDS and pairwise-likelihood maximization. 

\smallskip

The article is organized as follows: Section~\ref{sec:backgroud} provides a brief background in univariate and spatial extreme value statistics, and presents a short overview of MDS methods. Section~\ref{sec:modelfitting} details our two  different methods for fitting max-stable models using MDS and discusses parameter estimation. In Section~\ref{sec:results}, we compare the obtained models to a more classical max-stable model fitted in a (longitude, latitude, elevation) space with anisotropy and space rotation. In particular, we compare the ability of the models to reproduce the observed non-stationary dependencies. Finally, we discuss the advantages and drawbacks of the proposed approaches. Further diagnostic plots and  some commented R code is provided in the supplementary material. 

\section{Background}
\label{sec:backgroud}
\subsection{Data}

The dataset under study is provided by the Swiss Federal Office for Meteorology and Climatology (MeteoSwiss) and consists of daily measurements of rainfall from January 1st 1961 to December 31st 2013. Out of the $963$ stations for which data are available, we restrict our study to the $219$ stations that have no missing data. From the $219$ time series of daily precipitation, we are interested in the blockwise maxima computed over the autumn season, i.e. from 21 September to 20 December of each year. We choose to focus on this season because daily precipitation is highest in summer and autumn in northern Switzerland \citep{Umbricht:2013,Giannakaki:2015} and in autumn in southern Switzerland \citep{RN1563}. The blockwise maxima computed over the autumn  season are henceforth referred to as yearly maxima. Note that the number of stations (219) in the dataset is larger than in the references cited in the introduction (at most approx. $100$ stations in \citet{blanchet2011spatial}). 

\subsection{Univariate extreme value theory}

Let $\Xset = \{ x_1,\hdots,x_n \}$ denote the set of all $n = 219$ stations and let $x\in\Xset$. The yearly maximum precipitation $Z(x)$ at station $x$ is expected to approximately have a Generalized Extreme-Value (GEV) distribution,  
\begin{equation}
F_x(u) := \exp\left( - \left( 1 + \xi(x) \frac{ u - \mu(x) }{ \sigma(x) }   \right)^{-1/\xi(x)}_+  \right),
\end{equation}
where $(\cdot)_+ := \max(0,\cdot)$ and $\mu(x), \sigma(x), \xi(x)$ are the location, scale,  and shape parameter of the GEV at station $x$, respectively. When looking at spatial dependencies, i.e. at the joint distribution of the random vector $(Z(x))_{x\in\Xset} = (Z(x_1),\hdots,Z(x_n))\trans$, it is common to first transform the marginal distributions of the vector in order to obtain a vector $(Z^\star(x))_{x\in\Xset}$ with unit Fr\'echet (i.e. $\text{GEV}(1,1,1)$) margins. Following, e.g., \citet{blanchet2011spatial,davison2012review}, this approach is used in this paper and is performed through the transformation $u\mapsto -1/\log(F_x(u))$. Before transforming the block-maxima, the parameters $\mu(x), \sigma(x), \xi(x)$ need to be estimated for all $n$ stations. This estimation is not the focus of this work and we thus rely on a simplified regionalization procedure inspired from \citet{asadi2018optimal}.  In this procedure, the parameters for a station $x_{t_0}\in\Xset$ are estimated using the data of this station and its $J$ nearest neighbours (according to the classical Euclidean distance in the longitude, latitude space) using log-linear models and independent likelihood. We follow the procedure of \citet{asadi2018optimal} to determine the optimal number of neighbours $J$, which depends on the station $x_{t_0}$. We also follow \citet{ragulina2017generalized} to fix the constraint $\xi(x) \in [0,0.15]$ for all stations. Finally, for $30$ of the $219$ stations, the regionalization model does not yield satisfactory qq-plots because of strong local variations in precipitation, e.g. in the Ticino region. For these stations, we instead use the classical maximum-likelihood fit, which is equivalent to choosing $J=0$.

\subsection{Max stable models}

Among all types of random processes, max-stable ones are the most suited to model block maxima  \citep{davison2012review} because they are  the only non-degenerate limit of pointwise maxima of i.i.d random processes \citep{de1984spectral}. Formally, a random process $(Y(x))_{x\in\Xset}$ is max-stable with unit Fr\'echet margins if the random process $(NY(x))_{x\in\Xset}$ has the same distribution as $(\max_{i=1,..,N} Y_i(x))_{x\in\Xset}$, where $N > 0$ and $Y_1,\hdots,Y_N$ are $N$ i.i.d copies of $Y$. A widely used representation is given in  \citet{de1984spectral}, where $(Z^\star(x))_{x\in\Xset}$ is represented as the pointwise maxima of infinitely many i.i.d random processes weighted by random coefficients:
\begin{equation}
\label{eq:representation}
Z^\star(x) := \max_{i\in\Nset} \eta_i W^\star_i(x),
\end{equation}
where the $\eta_i$'s are drawn from a Poisson process on $\Rset_+$ with intensity $z^{-2} \rm{d}z$  and $W^\star_1,W^\star_2,\hdots$ are i.i.d copies of a non-negative random process $W^\star$ satisfying $\esp W^\star(x) = 1$ for all $x\in\Xset$. Several choices of processes $W^\star$ can lead to different popular models. In this work, we focus on the geometric Gaussian process, i.e. $W^\star = \exp(\sigma W - \sigma^2/2 )$ where $\sigma>0$ and $W$ is a centered Gaussian process with unit pointwise variance.  Although this choice is a particular case of the larger class of so-called Brown-Resnick processes, this model will be referred to as the Brown-Resnick model. 

\subsection{Extremal coefficients}
\label{sec:Extremalcoefficients}

For a max-stable process $Z^\star$ with unit Fr\'echet margins and for an arbitrary set of locations $\mathcal{D}_{1:N} := (x_1,\hdots,x_N)$, the finite-dimensional cumulative distributions of $Z^\star$ can be written as
\begin{equation}
\mathbb{P}(Z^\star(x_1) \leq z_1,\hdots, 
Z^\star(x_N) \leq z_N
) = \exp( -V_{1:N}(z_1,\hdots,z_N) ),
\label{cdf:general}
\end{equation} 
for some function $V_{1:N}:(z_1,\hdots,z_N)\in\Rset^N \mapsto \Rset$ called the exponent function \citep{resnick1987}. If we set $z_1 = \hdots = z_N := z$, Equation~\eqref{cdf:general} can be rewritten as 
\begin{align*}
\mathbb{P}(Z^\star(x_1) \leq z,\hdots, 
Z^\star(x_N) \leq z
) =& \exp( -V_{1:N}(1,\hdots,1)/z ) \\
:=& \exp( - \theta_{1:N} / z), 
\end{align*}
where $\theta_{1:N}$ is the so-called extremal coefficient associated with the set of locations $\mathcal{D}_{1:N}$. Practitioners are often interested in the pairwise extremal coefficients $\theta_{ij}$ between stations $x_i$ and $x_j$. For a Brown-Resnick model with parameter $\sigma>0$, this extremal coefficient is given by
\begin{align}
\label{eq:cov_to_theta_BR}
\theta_{ij} =& \ 
2 \Phi \left(\sqrt{ \frac{\sigma^2}{2} (1-k(x_i,x_j))} \right), 
\end{align}
where $k$ is the covariance function of the centered Gaussian process $W$ and $\Phi$ is the cumulative distribution function (c.d.f.) of the standard normal distribution. Throughout this paper, all covariances $k(x_i,x_j)$ are also correlations, since $W$ has unit  variance. One may note from Equation~\eqref{eq:cov_to_theta_BR} that if $k$ is a stationary covariance function then the modeled extremal coefficient depends directly on the (possibly anisotropic) distance between the stations. 

\subsection{Pairwise likelihood maximization and climate space}

Let $p$ denote the number of years of data, $n$ the number of stations, and $z^\star_{ik}$ the rescaled maximum precipitation at station $x_i$ during year $k$. The pairwise log-likelihood of some max-stable model with parameters $\gamma$ is
\begin{equation}
\label{pairwise:likelihood:gamma}
\ell(\gamma) = \sum_{1\leq i < j \leq n} \sum_{k=1}^p 
\log f_{ij}( z^{\star}_{ik} , z^{\star}_{jk};  \gamma) := \sum_{1\leq i < j \leq n} \ell_{ij}
\end{equation}
where $f_{ij}$ is the bivariate density of the random vector $(Z^\star(x_i),Z^\star(x_j))$ and $Z^\star$ is a max-stable process with unit Fr\'echet margins. In the particular case of Brown-Resnick models with parameter $\sigma$ and covariance   function $k$, $f_{ij}$ is obtained by differentiating the following bivariate c.d.f.:
\begin{align}
F_{ij}(z_i,z_j) =& \exp\left[- \left[
\frac{1}{z_i} \Phi\left( \nu_{ij} + \frac{\log(z_j/z_i)}{2\nu_{ij}}  \right) 
+ \frac{1}{z_j} \Phi\left( \nu_{ij} + \frac{\log(z_i/z_j)}{2\nu_{ij}}  \right) 
\right]
\right], 
\label{eq:F-BR}
\end{align}
where $\nu_{ij} = \sqrt{ \sigma^2 (1 - k(x_i,x_j))/2  }$. The latter equation shows that the contribution $\ell_{ij}$ of the pair $(i,j)$ of stations to the pairwise likelihood of Equation~\eqref{pairwise:likelihood:gamma} is directly linked to the covariance $k(x_i,x_j)$ and hence to the parameters of the covariance functions and other parameters accounting for space transformation. An example is given in \citet{blanchet2011spatial} where $\gamma$ is a set of parameters that allows space transformation. Considering the space of spatial coordinates (longitude, latitude, elevation), space transformation can be performed by working with spatial coordinates left-multiplied by the matrix
\begin{equation}
\label{eq:matrixV}
U = \left( 
\begin{array}{ccc}
c_1 \cos(\beta) & -c_1 \sin(\beta) & 0 \\ 
c_2 \sin(\beta) & c_2 \cos(\beta) & 0 \\ 
0 & 0 & c_3
\end{array} 
\right)
\end{equation}
where $\beta$ is a space rotation parameter and $c_1,c_2,c_3$ are parameters to account for anisotropy. Other covariates can be taken into account, and  additional anisotropy parameters $c_4,c_5,\hdots$ can be added to the diagonal of $U$. In \citet{blanchet2011spatial}, the space of spatial coordinates and covariates transformed with the matrix $U$ is called ``climate space''. The use of a climate space might be seen as a well-chosen transformation of the spatial coordinates of the stations, with the goal of yielding covariances $k(x_i,x_j)$ that lead to a large pairwise likelihood. If a stationary covariance function is used, the latter boils down to tuning the station interdistances.  This interpretation motivates the idea of using multidimensional scaling in lieu of a parametric space transformation.

\subsection{Multidimensional scaling}
\label{sec:MDS}

Let $D$ be a given $n\times n$ dissimilarity matrix and let $d>0$ be an integer. 
Multidimensional scaling (MDS) techniques aim at finding a configuration of $n$ points in $\Rset^d$ in such a way that the obtained $n\times n$ interpoint Euclidean distance matrix is as close as possible to $D$, with respect to some stress function. In many applications, MDS is used to provide graphical displays of $n$ points, which ease the interpretation of an $n\times n$ distance or dissimilarity matrix $D$ \citep{borg2005modern}. For visualization, the dimension $d$ is thus often set to $2$ or $3$, although the algorithms usually remain valid for larger values of $d$. Different choices of stress functions, as well as different algorithms to minimize the stress, yield a large set of MDS techniques. 
Below, we provide a list of the ones that are considered in this work.

\subsubsection{Classical scaling}
\label{sec:Classicalscaling}
Let $D^2\in\Rset^{n\times n}$ be a squared \textit{Euclidean} distance matrix; i.e. a matrix such that there exists a configuration of $n$ points in $\Rset^{n-1}$ (or in a lower-dimensional space) with a squared interpoint distance matrix \textit{exactly} equal to  $D^2$. Classical scaling follows the work of \citet{torgerson1952,torgerson1958} and a complete presentation and bibliography is given in \citet{borg2005modern}, Chapter 12. We define
\begin{align*}
J :=& I - \frac{1}{n} \mathbbm{1} \mathbbm{1}\trans \\
B :=& -\frac{1}{2} J D^2 J,
\end{align*}
where $I$ is the identity matrix and $\mathbbm{1}$ is the column vector of size $n$ with elements all equal to one. One can show that $B$ is the Gram matrix $B=X X\trans$ of a set of n points with coordinates $X\in\Rset^{n\times (n-1)}$ and squared distance matrix $D^2$. The coordinates $X$ are thus obtained by factoring $B = Q \Lambda Q\trans$, where $\Lambda$ is a diagonal matrix with sorted eigenvalues (in descending order), and by setting $X = Q \Lambda^{1/2}$. If the dimension $d$ is set to $n-1$, one obtains $n$ points in $\Rset^{n-1}$ with a squared distance exactly equal to $D^2$. If $d$ is less than $n-1$, one can take the first $d$ columns of $Q \Lambda^{1/2}$ -- which we denote as $X_d$ -- to obtain a set of points with squared distances \textit{close} to $D^2$. When $d < n-1$, one can show that the  solution $X_d$ is optimal if the stress function is defined as follows: 
\begin{equation}
L(X_d) := \left\Vert X_d X_d\trans  + \frac{1}{2} J D^2 J \right\Vert^2,
\end{equation} 
where $\Vert \cdot \Vert$ is the Frobenius norm in $\Rset^{n\times n}$. Classical scaling thus has two important drawbacks. First, the stress function given above is rather unnatural, and more intuitive stress functions will be presented below. Second, when the target dissimilarity matrix $D$ is not Euclidean, classical scaling is known to perform poorly \citep{cayton2006robust} because $B$ is not necessarily  positive-semidefinite. 

\subsubsection{Stress functions}
Let us consider a target dissimilarity matrix $D$ and a given MDS algorithm that yields a configuration $X_d$  of $n$ points in $\Rset^d$.  Let $d_{ij}(X_d)$ be the  distance between points $i$ and $j$ in this space. A natural stress function to assess the quality of the MDS algorithm is the raw stress function
\begin{equation}
\label{rawstress}
\sigma_r(X_d) = \sum_{1\leq i<j \leq n} (d_{ij}(X_d) - D_{ij})^2,
\end{equation}
where $D_{ij}$ is the element $i,j$ of the target dissimilarity matrix $D$. The term $i,j$ of the sum above can also be weighted with arbitrary weights $w_{ij}$. Many other choices of stress functions are possible  (see, \citet{borg2005modern}, Chapter 11, for a review). In this work, we considered the stress function of \citet{sammon1969nonlinear}, which is, up to a multiplicative  factor, given by
\begin{equation}
\label{sammonstress}
\sigma_S(X_d) = \sum_{1\leq i<j \leq n} \frac{(d_{ij}(X_d) - D_{ij})^2}{D_{ij}},
\end{equation}
and corresponds to the raw stress using weights $w_{ij} = 1/D_{ij}$. 
Let us also  mention the Stress-1 function which is used in \citet{sampsonguttorp1992JASA}
\begin{equation}
\sigma_1^2(X_d) =  \frac{\sum_{i<j} (d_{ij}(X_d) - \delta(D_{ij}))^2}{\sum_{i<j} d^2_{ij}(X_d)}
\end{equation}
where $\delta(\cdot)$ is a monotonic transformation of the target dissimilarities that is optimized together with the best configuration $X_d$. This so-called non-metric MDS is introduced by \citet{kruskal1964non}.  In \citet{sampsonguttorp1992JASA}, the inverse function $\delta^{-1}$ plays the role of a variogram function since the dissimilarities $D_{ij}$ are variances  of the difference $Y_i - Y_j$ of the spatial process values at locations $i,j$. The algorithm thus delivers a configuration $X_d$ such that $\delta^{-1}(d_{ij}(X_d)) \approx D_{ij}$. Our use of MDS shares similarities with that of \citet{sampsonguttorp1992JASA} because for our application the function $\delta^{-1}$ will transform some ``ideal covariances'' between stations (as detailed in Section~\ref{sec:idealcovariancematrix}) into  distances. However, we will not need to use the non-metric MDS algorithm of \citet{kruskal1964non} because only a finite number of functions $\delta$ are tested.

\subsubsection{Minimizing the stress functions}

Many possible methods exist to minimize the stress functions presented previously and a review is given in \citet{borg2005modern}. Here, we only mention the methods considered in this work. Many techniques belong to the family of gradient descent minimization  algorithms. For example,  \citet{sammon1969nonlinear} computes the $nd$-dimensional gradient of $\sigma_S$ given in Equation~\eqref{sammonstress} and uses iterative gradient-descent with an initial configuration obtained by classical MDS. The latter approach benefits from the convexity of the stress function. We also mention the so-called majorization algorithms, particularly the SMACOF algorithm \citep{smacof}, which aims at minimizing the raw stress $\sigma_r$.

\section{Proposed model fitting approach}
\label{sec:modelfitting}

We now detail our model-fitting procedures. Two methods are suggested, both of which rely on MDS. These methods will be referred to as ``method 1'' and ``method 2''. Later on, the methods are compared with the more classical approach based on climate spaces. The main difference between our two procedures lies in the computation of the so-called ideal covariance matrix, as detailed below.

\subsection{Ideal covariance matrix}
\label{sec:idealcovariancematrix}
\subsubsection{Method 1, based on fitting the estimated pairwise extremal coefficients}
Recall that $\Xset = (x_1,\hdots,x_n)$ is the set of all n = 219 stations. From our $53$ years of data, it is possible to estimate a $n\times n$ matrix of pairwise extremal coefficients $\hat{\theta}$. We choose to estimate these using the F-madogram estimator of \citet{cooley2006variograms}. Let us now consider a Brown-Resnick process with parameter $\sigma$.  Equation~\eqref{eq:cov_to_theta_BR} links the extremal coefficient between stations $x_i,x_j$ to the covariance, or correlation $k(x_i,x_j)$, and can be inverted, meaning that it is possible to compute a covariance, denoted by $k^{(1)}_{i,j,\text{ideal}}$, that leads to the extremal coefficient $\hat{\theta}_{ij}$ estimated from the data:
\begin{align}
\label{theta_to_k_br}
k^{(1)}_{i,j,\text{ideal}} :=& 
1 - \frac{2}{\sigma^2} \left(  \Phi^{-1} 
\left(\frac{\hat{\theta}_{ij}}{2}\right) \right)^2, 
\end{align} 
where $\Phi^{-1}$ denotes the quantile function of the standard normal distribution and where we use the convention $\Phi^{-1}(1) = \infty$. Since we will work with covariance functions delivering strictly positive values, we construct the ideal covariance matrix $K^{(1)}$ by flooring the ideal pairwise covariances with a minimum value $\varepsilon > 0$: 
\begin{align}
K^{(1)}_{ij} = \max(\varepsilon, k^{(1)}_{i,j,\text{ideal}}).
\end{align} 
The value of $\varepsilon$ could be a parameter of our actual model, but for simplicity we fix it at $\exp(-3)\approx 0.05$. Note that the matrix $K^{(1)}$ is not a covariance matrix because it has no reason to be positive definite. Also, $K^{(1)}$ depends on $\sigma$. The main idea with $K^{(1)}$ is that, if we find some Gaussian process, some $d$-dimensional latent space, and a well-chosen placement of our $n$ stations in this space, then the covariance matrix at our stations might be ``close'' to $K^{(1)}$. Hence, our max-stable process will be able to reproduce well the extremal coefficients estimated from the data. Note  that the idea of fitting the estimated extremal coefficients was already used in \citet{smith1990}. In the rest of this paper, we will be interested in the extremal coefficients  mean-squared error (MSE), or misfit, defined here as 
\begin{equation}
\label{eq:thetaMSE}
\theta^\text{MSE} = \frac{1}{n^2}\sum_{1\leq i,j \leq n} (\theta_{ij} - \hat{\theta}_{ij})^2.
\end{equation}
where $\theta_{ij}$ is the modeled extremal coefficient between stations $x_i$ and $x_j$.

\subsubsection{Method 2, based on pairwise likelihood maximization}
\begin{figure}[!htbp]
\centering
\includegraphics[width=0.65\textwidth]
{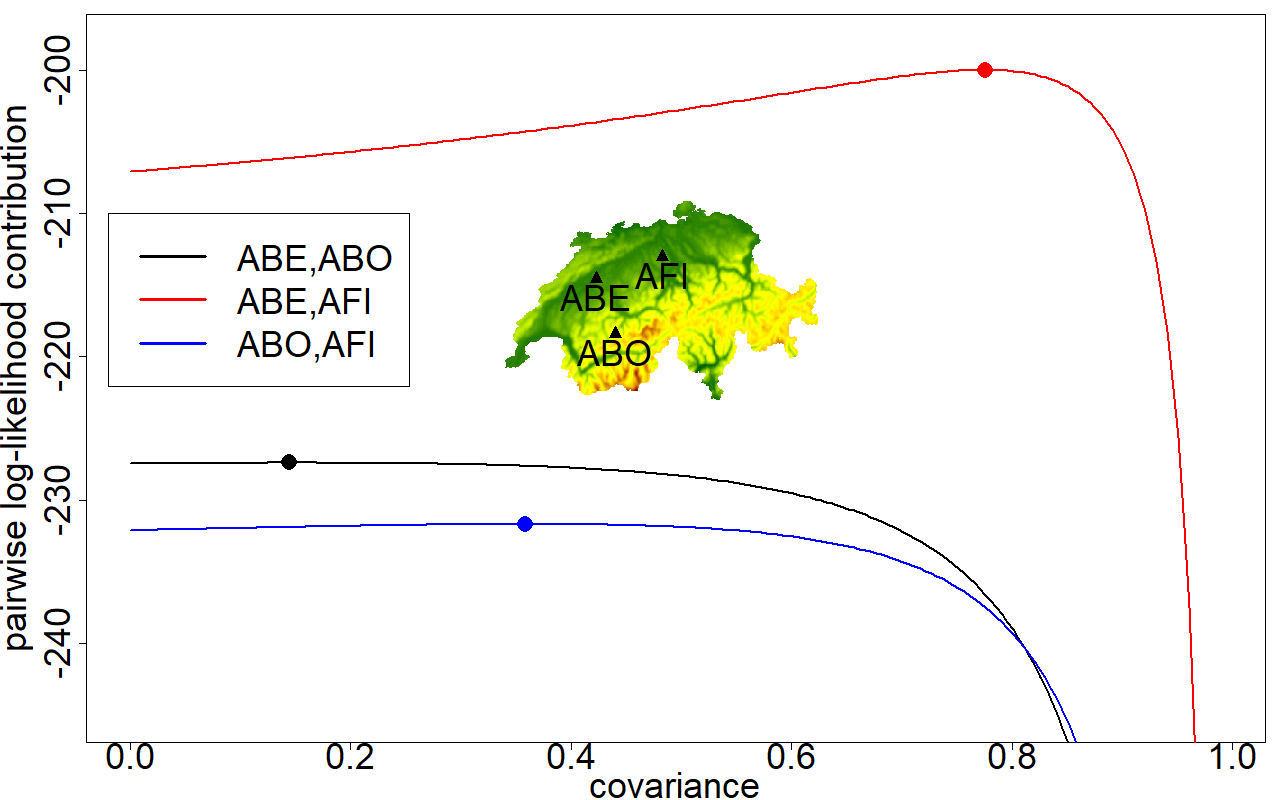}
\caption
{
Contribution $\ell_{ij}$ of $3$ different pairs of stations to the pairwise log-likelihood as a function of their covariance $k(x_i,x_j)$. The locations of the 3 stations forming the 3 pairs are indicated on the map. For each pair of stations, the covariance maximizing $\ell_{ij}$ is indicated with a dot. Since the covariances are positive and are also correlations, they are in the interval $[0,1]$.
}
\label{fig:PairwiseLikContribution_Suisse}
\end{figure}

Let us consider again a Brown-Resnick process with parameter $\sigma$ and a pair of stations $(x_i,x_j)$. The contribution of the Fr\'echet-transformed data from this pair of stations to the pairwise log-likelihood is given by the term $\ell_{ij}$ in Equation~\eqref{pairwise:likelihood:gamma} and depends on the covariance or correlation  $k(x_i,x_j)$ (see Equation~\eqref{eq:F-BR} and the expression of $\nu_{ij}$). When $\sigma$ is fixed, it is possible to plot $\ell_{ij}$ values as a function of $k(x_i,x_j)$ and see where $\ell_{ij}$ is maximized. An example is given in Figure~\ref{fig:PairwiseLikContribution_Suisse} where $\ell_{ij}$ is plotted for 3 different pairs of stations $(x_i,x_j)$. 
For each pair of stations, there exists an ideal covariance $k^{(2)}_{i,j,\text{ideal}}$ that  maximizes $\ell_{ij}$. In our work, the search for the ideal covariance is performed on the interval $[0,0.99]$ to avoid the numerical issues caused by correlations of $1$. The ideal covariance matrix $K^{(2)}$ is obtained by flooring the $k^{(2)}_{i,j,\text{ideal}}$'s with the same minimum value $\varepsilon > 0$ as in the previous method. Again, notice that $K^{(2)}$ depends on $\sigma$. The choice of this parameter will be discussed in Section~\ref{Parameterselection}.

\subsection{Spatial max-stable model from the ideal covariance matrix}
\subsubsection{Ideal distance matrix and MDS}
Let us consider a centered Gaussian process $W$ with unit pointwise variance and covariance (or correlation) function $k$. We assume that $k$ is stationary, strictly decreasing, strictly positive, and with infinite support, i.e. $k(0) = 1$ and $k(h)$ goes to $0$ as $h$ goes to infinity. With these conditions the inverse function $k^{-1}$ is well defined and goes from $(0,1]$ to $\Rset$. Let $K^\star$ be the ideal covariance matrix obtained using the covariance function $k$ and one of the two methods described above, assuming that we have fixed the parameter $\sigma$ of the Brown-Resnick model. The ideal covariance matrix is linked to an ideal station interdistance matrix $D^\star$ as follows:
\begin{equation}
D^\star := k^{-1}(K^\star).
\end{equation}
The previous equation shows that, if there exist some latent space in which the station interdistance matrix is $D^\star$, then the use of the Gaussian process $W$ in our Brown-Resnick model with parameter $\sigma$ will yield the covariance matrix $K^\star$ and hence will yield a good extremal coefficients fit (method 1) or a large pairwise likelihood (method 2), depending on how $K^\star$ is constructed. The question thus boils down to finding a latent space in which the station's distance matrix is $D^\star$.  Since $D^\star$ has no reason to be an Euclidean distance matrix, such latent space does not always exist. However, it might be possible to obtain distance matrices that are \textit{close} to $D^\star$, and this is where MDS techniques are used. In this work, we use MDS algorithms on the matrix $D^\star$ to obtain a placement of our $n$ stations in a latent space of fixed dimension $d$. Among the different MDS techniques available (see Section~\ref{sec:MDS}), we use the algorithm of \cite{sammon1969nonlinear},  which uses an $nd$-dimensional gradient descent to minimize the stress function of Equation~\eqref{sammonstress}. For the sake of conciseness, results with competing MDS techniques are not presented. Experiments suggested that Sammon's algorithm constantly yields a better pairwise extremal coefficients fit (method 1) and a better pairwise likelihood (method 2). 

\subsubsection{Getting a real spatial model}
\label{sec:realspatialmodel}
\begin{figure}[!htbp]
\centering
\includegraphics[width=0.65\textwidth]
{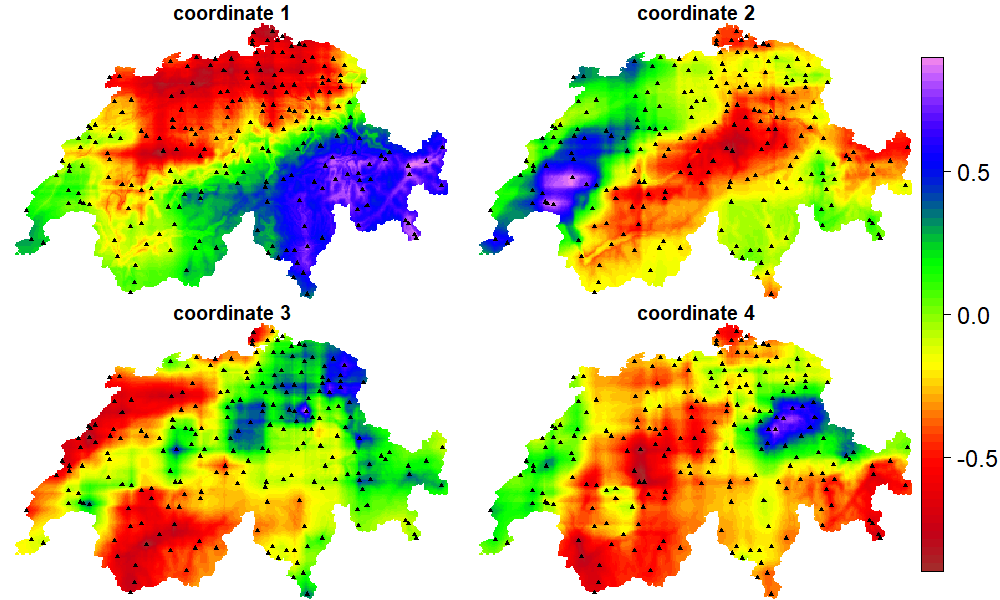}
\caption
{
Ordinary Kriging predictors $\psi_1,\hdots,\psi_4$ obtained by interpolating the MDS map. Here, $K^\star$ is built using method 1 and $\sigma = 2.8$. The ideal distance matrix $D^\star = k^{-1}(K^\star)$ is obtained using a stationary isotropic power-exponential covariance function $k$ with exponent $\alpha = 2$. The $4$ Ordinary Kriging models $\psi_1,\hdots,\psi_4$ use an anisotropic exponential covariance function with parameters estimated by maximum likelihood. 
}
\label{fig:coord4d}
\end{figure}

When the parameters of the proposed model (i.e., the latent space dimension $d$, the covariance function $k$ and the Brown-Resnick parameter $\sigma$) are fixed, the use of an MDS method to warp the stations in the $d$-dimensional latent space  does not immediately yield a \textit{spatial} model in the latent space. Indeed, the MDS mapping that warps the stations in the latent space is not known explicitly everywhere, since it is known only at the station locations. Here, following \citet{sampsonguttorp1992JASA} and \citet{borg2005modern}, this mapping will be explicitly constructed everywhere using interpolation. Let $\bold{x_1},\hdots,\bold{x_n}$ be the coordinates of the stations $x_1,\hdots,x_n$ in the $3$-dimensional space (longitude, latitude, elevation). Let $\bold{y_1},\hdots,\bold{y_n}$ be the coordinates of these stations in the latent $d$-dimensional space, obtained using MDS.  Also, for all $i$, we use the notation $\bold{y_i} = (y_{i,1},\hdots,y_{i,d})$. Interpolation of the MDS mapping can be simply performed by constructing $d$ Ordinary Kriging models (see \citet{ginsbourger:dicekrigingoptim} for a short description) for each of the $d$ coordinates in the latent space. We denote by $\psi_1,\hdots,\psi_d$ the $d$ Ordinary Kriging predictors which satisfy $\psi_j(\bold{x_i}) = y_{i,j}$ for $1\leq j\leq d$ and $1\leq i\leq n$. In this work, these predictors are computed using the DiceKriging R package \citep{ginsbourger:dicekrigingoptim} and anisotropic exponential covariance functions, with parameters estimated by maximum likelihood. An example with $d = 4$ is given in Figure~\ref{fig:coord4d}. The $d$ kriging predictors, obtained by interpolating the MDS mapping, enable us to warp any location $\bold{x}$ in Switzerland to the $d$-dimensional latent space  by simply computing $\psi(\bold{x}) := (\psi_1(\bold{x}),\hdots,\psi_d(\bold{x}))$. The modeled covariance between two arbitrary locations $\bold{x}$ and $\bold{x^\prime}$ in Switzerland is thus simply given by $k(\psi(\bold{x}), \psi(\bold{x}^\prime))$, where $k$ is our covariance function in the latent space.

\subsection{Parameter selection}
\label{Parameterselection}

In the previous section, MDS is performed assuming that the covariance function $k$ of the Gaussian process $W$ is known, and assuming that the Brown-Resnick parameter $\sigma$ and the latent space dimension $d$ are known. In practice, all of these parameters need to be estimated. Parameter selection can be divided into two steps. The first step, described in Section~\ref{sec:choicecovariance}, consists of estimating a covariance function $k$ and its parameters for all the possible choices of $d$ and $\sigma$. In practice, a finite grid of $(d,\sigma)$ values is considered and the covariance function  $k$ is optimally chosen for all these $(d,\sigma)$ pairs. The second step is also sketched in Section~\ref{sec:choicecovariance} and consists in optimally choosing $\sigma$ when $d$ is fixed. Finally, the choice of $d$ is detailed in Section~\ref{sec:choiced}. 

\subsubsection{Choice of the parametric covariance function $k$ and of $\sigma$ when $d$ is fixed}
\label{sec:choicecovariance}

We first discuss the choice of the covariance function and, when applicable, of the parameters of this function. For our MDS application, among the family of  stationary covariance functions, it is sufficient to consider only isotropic ones. Indeed, the use of range parameters (sometimes called correlation lengths), that would account for anisotropy in the latent space, would simply yield a configuration of points $X_d$ with coordinates divided by the corresponding range parameter. In other words, all the models that we can produce with anisotropic covariance functions $k$ can also be obtained with isotropic ones. 

\smallskip

The considered covariance functions are  described in Table~\ref{table:covfunctions}. All these covariances can be used in spaces of arbitrary dimension $d$, a property that is mandatory for the application of our method. The computation of $k^{-1}(\cdot)$ is performed either analytically or numerically. Many covariance functions (e.g., spherical, circular, cubic, Gneiting) are omitted here because they cannot be used in arbitrary dimensions. 
\begin{table*}
\begin{tabular}{cccc}
\hline
Covariance 
name & \multicolumn{1}{c}{$k(h)$} & \multicolumn{1}{c}{$k^{-1}(c)$} & \multicolumn{1}{c}{Parameters} 
\\
\hline
Power Exponential & $\exp(-h^\alpha)$ & 
 $(-\log(c))^{1/\alpha}$ & $\alpha\in (0,2]$  \\
Mat\'ern($\nu = 3/2$) &    $(1+h) \exp(-h)$ 
& numerical computation & None  \\
Mat\'ern($\nu = 5/2$) &    $\left(1+h + \frac{h^2}{3}\right) \exp(-h)$ 
& numerical computation & None  \\
\hline
\end{tabular}
\caption{
Covariance functions of the centered Gaussian process $W$ for the considered Brown-Resnick model. All the considered covariance functions are isotropic and can be used in a space of arbitrary dimension $d > 0$. 
}
\label{table:covfunctions}
\end{table*}

\smallskip

\begin{figure}[!htbp]
\centering
\includegraphics[width=0.65\textwidth]
{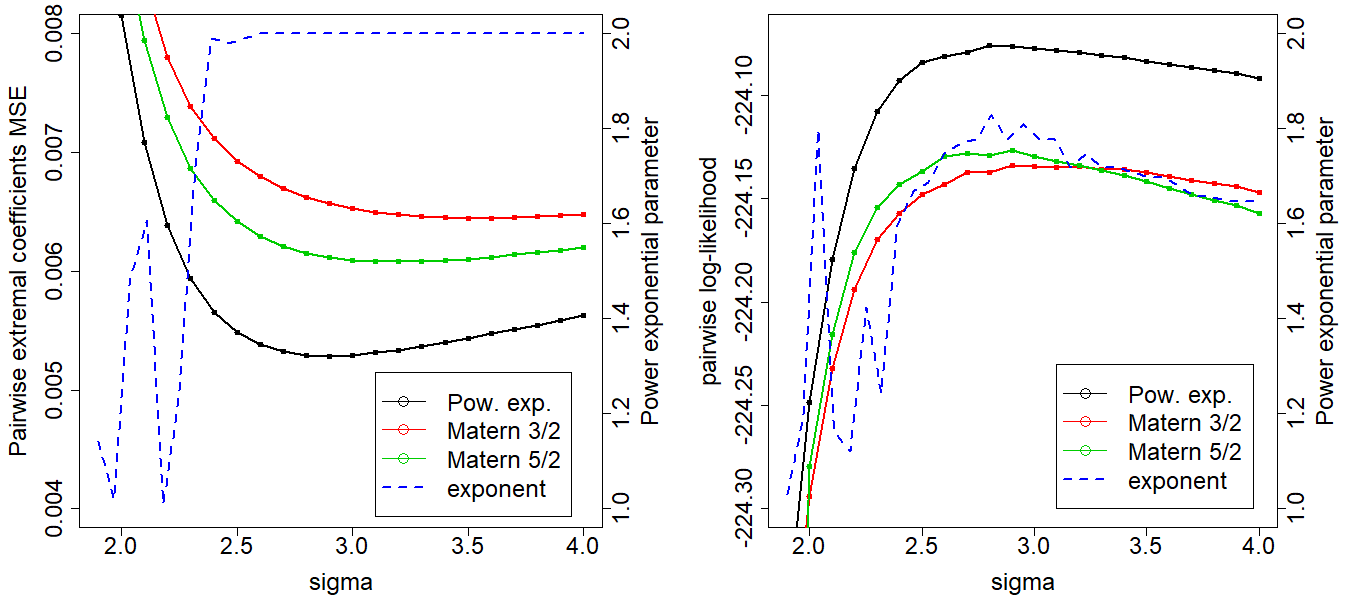}
\caption
{
Left plot, left y-axis: Extremal coefficient MSE, as a function of the parameter $\sigma$, obtained using method 1, $d=5$, and different covariance functions. 
Right y-axis: Parameter $\alpha$ of the power-exponential covariance function that minimizes $\theta^{\text{MSE}}$. 
Right plot: Pairwise log-likelihoods, as a function of the parameter $\sigma$,  
obtained using method 2, $d=5$, and different covariance functions. Right y-axis: Parameter $\alpha$ of the power-exponential covariance function that maximizes the pairwise likelihood.
}
\label{fig:covarianceparameter_choice}
\end{figure}

We first assume that $d$ is fixed. Through the use of MDS, a given choice of covariance function $k$ and of $\sigma$ is linked to a given pairwise extremal coefficients mean-squared error (method 1, see Equation~\eqref{eq:thetaMSE}) or a given pairwise log-likelihood (method 2). Figure~\ref{fig:covarianceparameter_choice}, left plot, sketches the choice of the covariance function and of $\sigma$ when method 1 is used. The extremal coefficient MSE (see  Equation~\eqref{eq:thetaMSE}) is shown as a function of $\sigma$, in the case where $d=5$, for all of the tested covariance functions. Since the power exponential function has a parameter, we also plot its value. The exponent $\alpha$ in the power-exponential covariance is chosen to minimize $\theta^{\text{MSE}}$ and depends on $\sigma$. In the right plot, method 2 is used and we show a pairwise-likelihood as a function of the covariance function choice and of $\sigma$. Here, when the power-exponential covariance is used, the exponent $\alpha$ is chosen to maximize the pairwise likelihood.
Figure~\ref{fig:covarianceparameter_choice} shows that, for each value of $\sigma$ and for both methods 1 and 2, the power exponential covariance yields better extremal coefficient misfits (left plot, method 1) or larger log-likelihoods (right plot, method 2) than the Mat\'ern(3/2) or Mat\'ern(5/2) covariance functions. This has been observed for all of the considered dimensions $d$ of the latent space. The result is not surprising because the power exponential covariance has a parameter, $\alpha$, that can be optimally tuned, in contrast to the Mat\'ern(5/2) or Mat\'ern(3/2) covariances. 

\smallskip

The Mat\'ern($\nu$) covariance, with varying smoothness $\nu$, might outperform the power exponential covariance. However, this has not been investigated further because computing and inverting the Mat\'ern($\nu$) covariance for arbitrary $\nu$ is too computer intensive. In conclusion we use the power-exponential covariance for all $(d,\sigma)$ values, with an exponent $\alpha$ that varies with $(d,\sigma)$. This is shown in Figure~\ref{fig:covarianceparameter_choice} for the particular case $d=5$. Figure~\ref{fig:covarianceparameter_choice} also suggests an optimal choice for $\sigma$. At fixed $d$, if method 1 is used, $\sigma$ and the exponent $\alpha(d,\sigma)$ are chosen to minimize $\theta^{\text{MSE}}$. If method 2 is used, $\sigma$ and the exponent $\alpha(d,\sigma)$ are chosen to maximize the pairwise likelihood.

\subsubsection{Choice of $d$}
\label{sec:choiced}

When $d$ is fixed, the Brown-Resnick parameter $\sigma(d)$ and the power-exponential exponent $\alpha(\sigma(d),d)$ are chosen by optimizing the pairwise extremal coefficients MSE or the pairwise log-likelihood, depending on which method is used. The first point regarding the choice of the latent space dimension $d$ is the range of possible $d$ values. In this work we choose $d$ in the set $\mathcal{S}:=\{2,3,4,5,6\}$. The choice of the maximum value $d=6$ of the latent space dimension is motivated by the results of \citet{perrin2003nonstationarity},  which indicate that a non-stationary field in space of dimension $v$ can be represented as second-order stationary in dimension $2v$. Applying this result to a non-stationary random field in the (longitude, latitude, elevation) space yields a maximum latent space dimension of $d=6$. To choose the latent space dimension $d$ in $\mathcal{S}$, we simply use the idea that incrementing the value of $d$ (that is: choosing the value $d+1$ instead of $d$) should substantially improve the obtained model. This means that if $d$ is incremented, then $\theta^{\text{MSE}}$ should decrease significantly enough (method 1) or the obtained pairwise likelihood should increase significantly enough (method 2). More specifically, if method 2 is used and if $\ell_{d}$ is the log-likelihood obtained from the best model with latent space dimension $d$, the dimension $d+1$ is chosen over the dimension $d$ if $\ell_{d+1} / \ell_{d} > r_2$, where $r_2$ is a corresponding minimal rate of improvement. Similarly, for method 1, if $\theta^\text{MSE}_{d}$ is the pairwise extremal coefficients MSE of the best model in dimension $d$, the dimension $d+1$ is chosen over the dimension $d$ if $1-\theta^\text{MSE}_{d+1}/\theta^\text{MSE}_{d} > r_1$. The rates $r_1,r_2$ could possibly be found by considering criteria involving the number of parameters of the model, like the Akaike Information Criterion. However, in our case, the number of parameters in our MDS-based model is not straightforward to determine. In this work, we plot the values of $\ell_{d}$ and $\theta^\text{MSE}_{d}$ for $d\in\mathcal{S}$ and visually choose $r_1= 5\%$, and $r_2=0.025\%$. This choice yields $d=5$ for both methods 1 and 2.

\section{Results and comparisons}
\label{sec:results}
\subsection{Introduction: competing models}

In this section, we compare the max-stable models obtained using the two proposed model fitting methods with a more classical model fitted in a climate space, following \citet{blanchet2011spatial}. The classical model uses the space (longitude, latitude, elevation) and a parametric space deformation performed with the  matrix $U$ of Equation~\eqref{eq:matrixV}. The parameters of the model are the Brown-Resnick parameter $\sigma$, the angle $\beta$ and the anisotropy parameters $c_1,c_2,c_3$. In addition, we use the isotropic power-exponential covariance function in the climate space, which has an exponent parameter $\alpha$. The $6$ parameters of the model are estimated by maximizing the pairwise-likelihood obtained with the data from all $n = 219$ stations. As in  \citet{blanchet2011spatial}, we maximize the pairwise likelihood using a simple line-search algorithm. The likelihood is sequentially maximized with respect to one parameter while keeping the other parameters fixed, and the procedure is iterated until convergence. For the classical model, we considered adding additional covariates to the climate space, like the mean precipitation. However, the additional covariates did not improve the  obtained likelihoods. The parameters of the fitted models, which are compared in the next sections, are summarized in  Table~\ref{table:competingmodels}.
\begin{table}
\begin{tabular}{c|c|c|c}
 & method 1 & method 2 & classical model \\ 
\hline 
latent/climate space dimension $d$ & $5$  & $5$  & $3$ \\
Brown-Resnick parameter $\sigma$ & 
$2.9$   & 
$2.8$   & 
$2.37$ \\
Power-exponential exponent $\alpha$ & 
$2$  & 
$1.72$ & 
$1.09$ \\
\end{tabular} 
\caption{
Description of the three fitted models that are compared. All of these models are Brown-Resnick models with an isotropic (in the latent or climate space) power-exponential covariance function with parameter $\alpha$. 
For the classical model, the estimated space-deformation parameters are $(c_1,c_2,c_3,\beta)\approx (0.70,1.52,\num{3.3e-4},-0.11)$, see Equation~\eqref{eq:matrixV}.
}
\label{table:competingmodels}
\end{table}
\subsection{Extremal coefficients fits and pairwise log-likelihoods}
\label{sec:results_fit_pairwiselik}

The fitted max-stable models should be able to reproduce the pairwise extremal coefficients  estimated from the data. They should further give good overall pairwise log-likelihoods. In our experiments, both methods 1 and 2 yield better pairwise extremal coefficients fit than the classical model, as well as better pairwise likelihoods. To determine the significance of these results and to challenge the proposed methods, we decided to perform $50$ additional random experiments. In each random experiment, $n_2$ test stations are not used to fit our models and thus the models are fitted based on only $n_1 := n - n_2$ training stations. The obtained extremal coefficient misfits and pairwise likelihoods are then obtained by using the MDS mapping for the $n_1$ training stations and by interpolating this mapping for the $n_2$ test stations (see Section~\ref{sec:realspatialmodel}). In the random experiments, $n_2$ is chosen uniformly in the set of integers between $25$ and $50$. The $n_2$ test stations are then chosen using the space-filling algorithm of the BalancedSampling R package \citep{balancedsampling,grafstrom2012spatially}. 

\smallskip

\begin{figure}[!htbp]
\centering
\includegraphics[width=0.65\textwidth]
{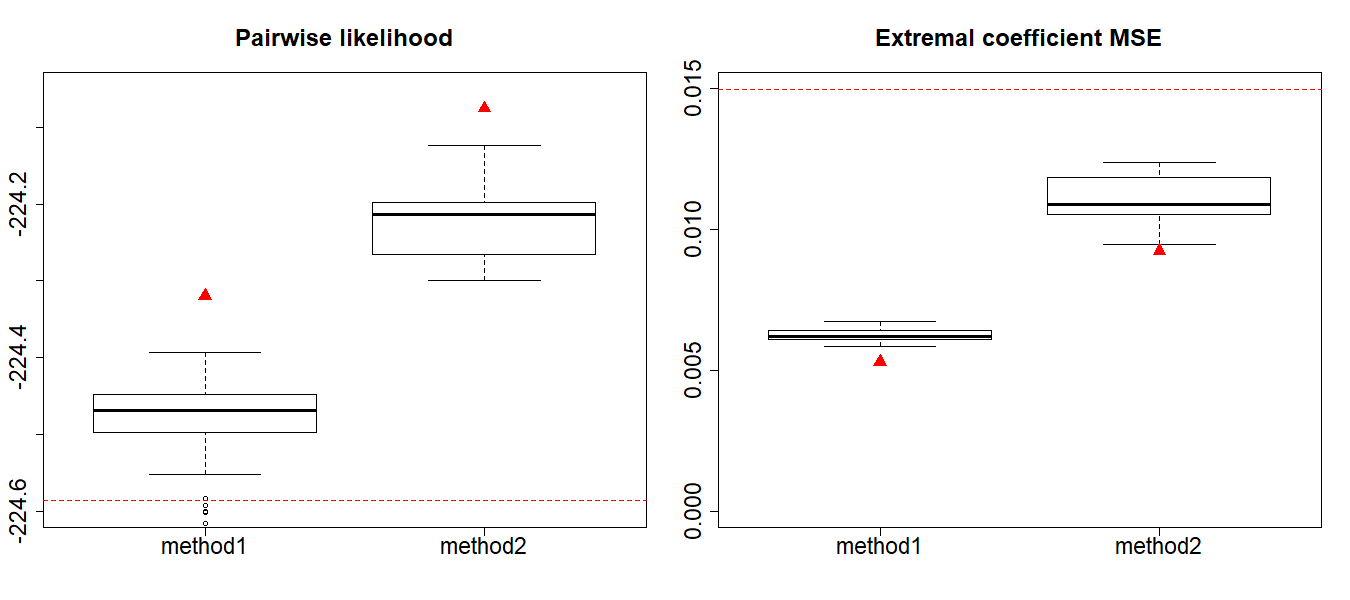}
\caption
{
Pairwise log-likelihoods (left plot) and $\theta^{\text{MSE}}$ (right plot) obtained using the proposed full models (red triangles) and the classical model (red dashed curve). The boxplots indicate the results of $50$ random experiments in which $n_2$ stations are randomly selected and are not used to fit the proposed models.
}
\label{fig:boxplot_results}
\end{figure}

Figure~\ref{fig:boxplot_results} gives the obtained pairwise log-likelihoods (left plot) and $\theta^\text{MSE}$ (right plot) for our full models (methods 1 and 2 using all $n=219$ stations), the classical model, and all $50$ random experiments. The red dashed curve indicates the pairwise likelihood or $\theta^{\text{MSE}}$ of the classical model. The red triangles indicate the pairwise likelihood or $\theta^{\text{MSE}}$ of the proposed full models, and the boxplots indicate the results of the $50$ random experiments. Unsurprisingly, removing $n_2$ test stations from the dataset yields models with a slightly decreased pairwise likelihood or extremal coefficient fit. The full model fitted with method 1 (optimizing the extremal coefficient fit) has  better pairwise likelihoods than the classical model, even though the fitting method does not aim to maximize this likelihood. In addition, approximately $90\%$ of the random models -- which use less data than the classical model -- have a better pairwise likelihood than the classical model. Compared to the classical model, method 1 manages to divide  $\theta^{\text{MSE}}$ by a factor $3$ for the full model, and by a factor larger than $2$ for all $50$ random experiments. The models fitted with method 2 (optimizing the pairwise likelihood) always have better pairwise likelihoods than the classical model and also always improve the misfit $\theta^{\text{MSE}}$. Unsurprisingly, method 1 still outperforms method 2 with respect to extremal coefficient fitting. We noticed, however, that for the $50$ random experiments of method 2, the extremal coefficient misfits are close for pairs of training stations (used to fit the model) and pairs of test stations (not used to fit the model), which shows that overfitting is avoided. 

\subsection{Extremal coefficients maps}
\label{sec:extrcoeffmap}

\begin{figure}[!htbp]
\centering
\includegraphics[width=0.65\textwidth]{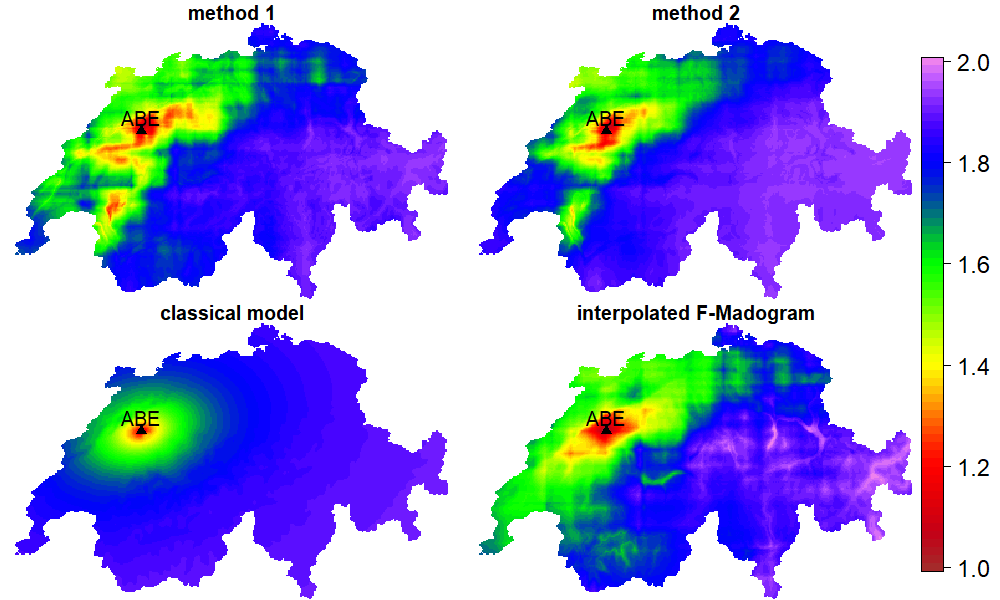}
\caption
{
Maps of pairwise extremal coefficients between the Aarberg station and other points in Switzerland. The bottom right map is obtained from the observed extremal coefficients (estimated with the F-madogram estimator) and with  kriging using an anisotropic exponential covariance function in the  (longitude, latitude, elevation) space.
}
\label{fig:extrcoeffmap_station}
\end{figure}

We show in this section that the main advantage of the proposed model fitting methods is their natural ability to handle complex non-stationary dependencies. Figure~\ref{fig:extrcoeffmap_station} shows a map of pairwise extremal coefficients  between a given station (Aarberg) and any other point in Switzerland. The first three maps are computed from the three competing models (method 1, method 2, classical model). The fourth map is computed from the extremal coefficients estimated from the data, using F-madogram. For visualization, the fourth map requires a spatial interpolation, which is performed here using Ordinary Kriging in the (longitude, latitude, elevation) space, with an anisotropic exponential covariance function. Even though the last map should be interpreted with care -- since it depends on how the interpolation is performed -- one clearly sees that methods 1 and 2 better  reproduce the strong non-stationarity observed in the data. For many stations in Switzerland, the map of extremal coefficients cannot be well reproduced with ellipses in the (longitude, latitude,  elevation) space, as assumed by the classical model.
The maps in Figure~\ref{fig:extrcoeffmap_station} can be produced for all $n=219$ stations,  and are released as supplementary material. Below, we produce these maps  for the $4$ stations after Aarberg, in alphabetical order. Notice that for the Adelboden station (``ABO'', i.e. the $4$ maps at the top left), method 1 is able to reproduce complex dependencies, where the set of locations with an extremal coefficient lower than 1.6 is not a connected set. The same phenomenon can be observed for many other stations. 

\smallskip

\begin{figure}[!htbp]
\centering
\includegraphics[width=0.495\textwidth]{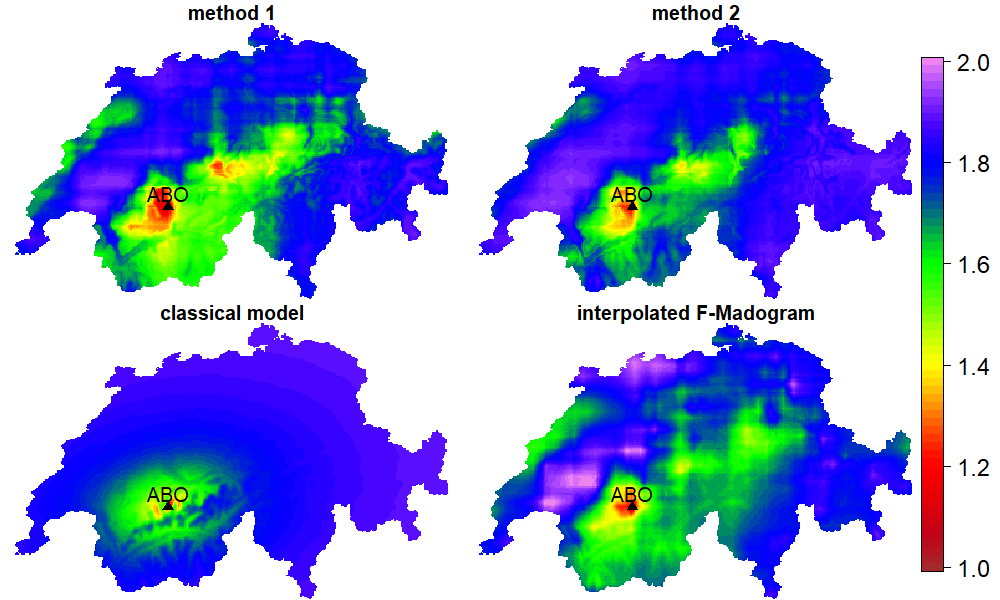}
\includegraphics[width=0.495\textwidth]{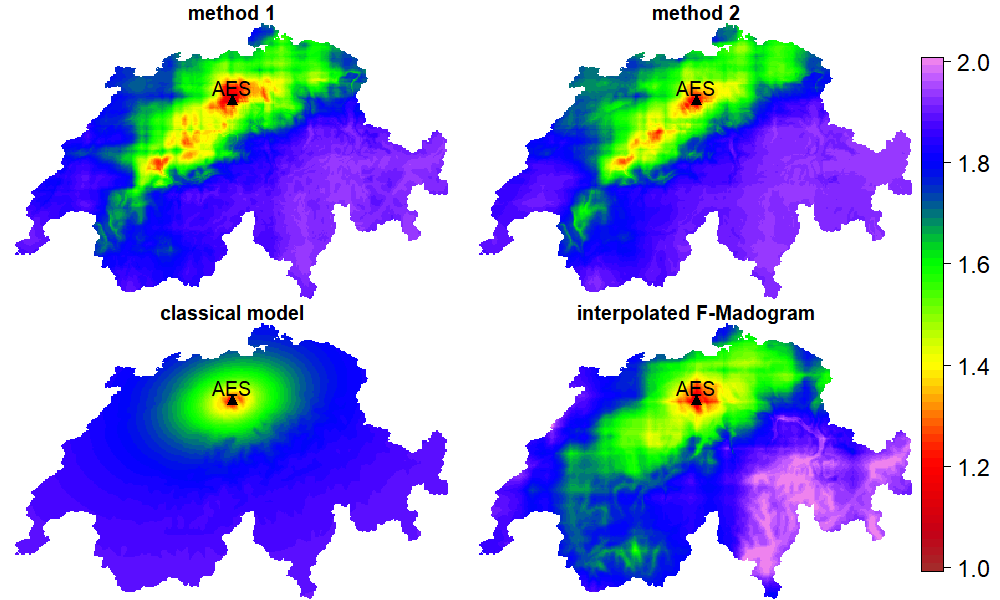}
\includegraphics[width=0.495\textwidth]{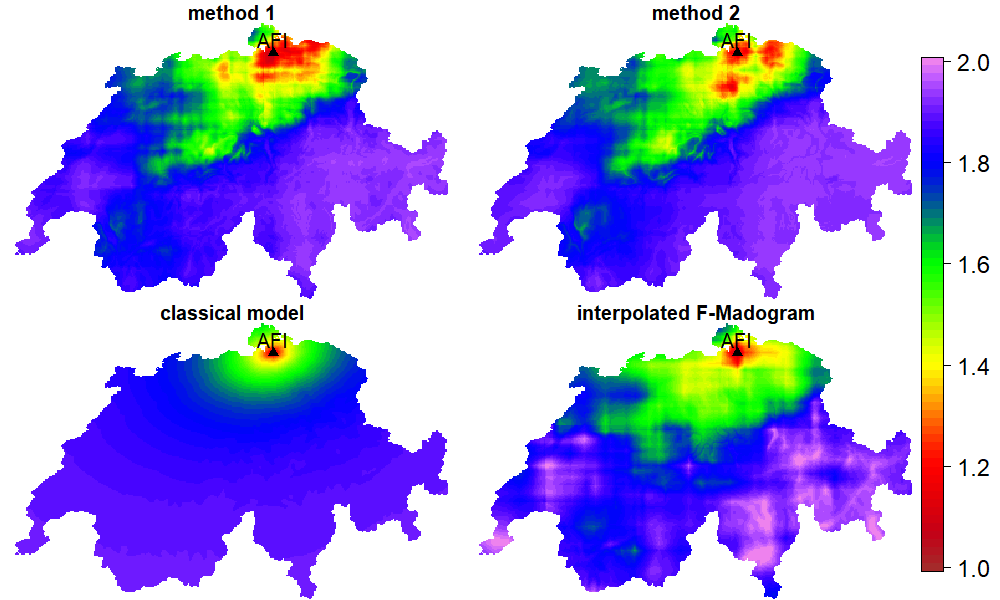}
\includegraphics[width=0.495\textwidth]{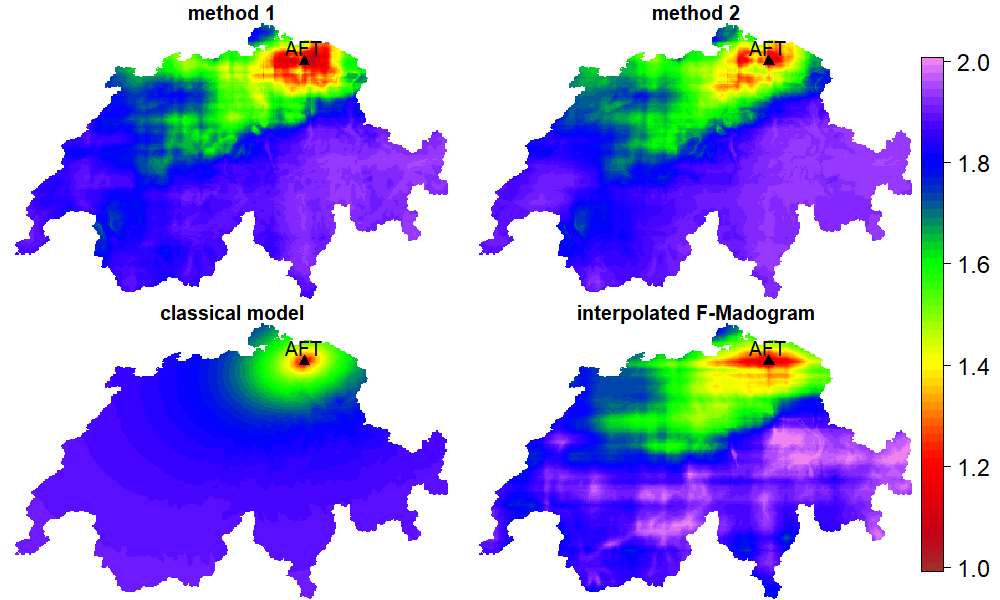}
\caption
{
Maps of pairwise extremal coefficients, similar to those in Figure~\ref{fig:extrcoeffmap_station}. The $4$ maps are plotted for the Adelboden (4 maps at the top left), Aesch (top right), Andelfingen (bottom left) and Affeltrangen (bottom right) stations.
}
\end{figure}

\subsection{Computation time}

The use of Sammon's MDS algorithm on matrices of size $n\times n$ is relatively fast.  With a standard computer with 3.40Ghz cpu and 8Gb of RAM, Sammon's algorithm takes a bit less than 0.1 second for the considered number of stations $n=219$. However, in the fitting procedure of Section~\ref{sec:modelfitting}, MDS is used a large number of times for different values of the parameters $\alpha,\sigma,d$. The initial step of our procedure is to build the ideal covariance matrix, which depends on $\sigma$ (see Section~\ref{sec:idealcovariancematrix}), but not on $d$ or $\alpha$. For method 1, this operation is very fast (see Equation~\eqref{theta_to_k_br}), but for method 2 it is not. Since the ideal covariance matrix depends on $\sigma$, we store all the ideal covariances matrices for method 2 for a grid of $\sigma$ values, ranging from $1$ to $4$ with a step of $0.1$. This initial step does not involve any MDS. Storing the $31$ possible ideal covariance matrices takes approximately $30$ minutes for method 2. For method 1, no storage is performed.

\smallskip

In the step detailed in Section~\ref{sec:choicecovariance}, we find the covariance parameters (here: the exponent $\alpha$ of the power-exponential covariance function) for all possible pairs $(d,\sigma)$. For each value of $d$, we have to run $31$  optimizations of an objective function that uses MDS. The objective function is the extremal coefficient MSE or the pairwise log-likelihood, depending on the method used. Here, the optimization with respect to $\alpha$ is simply performed on a grid of size $100$.  At fixed $d$, finding $\sigma$ and $\alpha$  requires $31\times 100 = 3'100$ calls to the MDS algorithm. Since this operation is repeated for $d=2,3,4,5,6$, the parameter estimation procedure (i.e. estimating $d,\sigma,\alpha$) takes approximately $25$ minutes for method 1, and $150$ minutes for method 2. Computation times for method 2 are higher because computing a pairwise likelihood is slower than computing an extremal coefficient MSE. For method 1, most of the computation time is due to   calls to Sammon's MDS algorithm.

\smallskip

Notice that in Section~\ref{sec:results_fit_pairwiselik} we perform $50$ random experiments by removing $n_2$ test stations from the dataset. To avoid multiplying the previous computation times by $50$, the mapping $(d,\sigma) \mapsto \alpha$ built for the full model is used for all $50$ experiments. Hence, at fixed $d$, optimizations are performed only with respect to $\sigma$, at the cost of $31$ calls to the MDS algorithm instead of $3'100$ calls.

\smallskip

In comparison, fitting the classical model requires the maximization of a pairwise likelihood with respect to six parameters, and takes approximately $4$ minutes. The improvement provided by the proposed models thus comes at the price of higher computational costs. The computational cost could be reduced by using faster procedures to optimize with respect to $\sigma$ and $\alpha$. In this work, we ran discrete optimizations on a grid of size $31\times 100$. Our methodology could be used with a larger number $n$ of stations since Sammon's MDS mapping is still relatively fast. The main limiting factor is the diagonalization of an $n\times n$ matrix, which is needed to obtain an initial placement of the stations through classical scaling, before using Sammon's gradient-descent.

\section{Conclusion and perspectives}
\label{sec:conclusion}

In this work, we introduce new max-stable model fitting procedures that rely on multidimensional scaling (MDS). Instead of relying on a stationary model in a climate space, obtained with some parametric space deformation of the (longitude, latitude, elevation) space, we use a stationary model in a latent space built with MDS. Compared to more classical approaches, the proposed methods are better able to reproduce non-stationary spatial dependencies, as shown by the extremal coefficient map in Figure~\ref{fig:extrcoeffmap_station}. This improved modeling of complex dependence structures comes at the price of increased computation time, especially for the method based on pairwise-likelihood maximization.

\smallskip

A possible future improvement is to implement an ad-hoc MDS algorithm where one could directly play with the coordinates $X_d$ of the $n$ stations in the latent space to directly maximize a log-likelihood or minimize a  pairwise extremal coefficients MSE. In that case, the notion of an ideal covariance matrix introduced in Section~\ref{sec:idealcovariancematrix} would become irrelevant. One of the main challenges for such a procedure would be its computation time, since MDS is performed many times to estimate covariance parameters, $\sigma$ and $d$.

\bigskip
\begin{center}
{\large\bf SUPPLEMENTAL MATERIALS}
\end{center}

\begin{description}

\item[SupplementaryMaterial:] Additional extremal coefficients maps (see Section~\ref{sec:extrcoeffmap}) are provided with comments. The supplementary material file contains a link to download the Fr\'echet-transformed dataset used in this work, as well as commented R code, allowing one to easily reproduce all of the results and figures in this article. A link for downloading all $219$ extremal coefficient maps is also provided. Finally additional diagnostic plots for the proposed models are given. (PDF file)

\end{description}

\bibliographystyle{jcgs}
\bibliography{technicalreport}

\end{document}